\documentstyle[12pt]{article}
\renewcommand
\baselinestretch{1.4}

\begin{document}

\hfill{UM-P-93/80}

\hfill{OZ-93/20}

\begin{center}
{\LARGE \bf Real Higgs singlet and the electroweak phase transition
in the standard model}

\vspace{4mm}

\renewcommand
\baselinestretch{0.8}\vspace{4mm}

{\sc J. Choi} and {\sc R. R. Volkas}\\

{\it
Research Centre for High Energy Physics,\\
School of Physics, University of Melbourne, \\
Parkville, Victoria 3052, Australia}

\renewcommand
\baselinestretch{1.4}

\vspace{5mm}
\end{center}

\begin{abstract}
The effective potential at finite temperature is constructed within the
minimal standard model when a real Higgs singlet is added on. We find
that a region of parameter space exists for which one can find a
first order transition strong enough to prevent the erasure due to
sphalerons of baryon asymmetry, while keeping the mass of the smallest
Higgs boson above the experimental lower bound of about 60 GeV.
\end{abstract}

\section{Introduction}
The possibility of baryogenesis at the electroweak phase transition
continues to provide an active area of theoretical work \cite{nels}.
Some topics in this area include: (i) constructing a more reliable effective
potential by taking the infrared divergences into account, (ii) finding
viable scenarios of bubble propagation at the electroweak phase
transition, and (iii) examining the various extensions of the
standard model (SM) that could possibly accommodate baryogenesis at
the electroweak scale.
This paper falls into the last category, in that we seek to find out
whether the minimal extension of the SM Higgs sector - adding only
a real singlet Higgs boson - can provide a first order phase transition
as necessary for creating a net baryon number. And if so, whether the
strength of this phase transition is sufficient so that a subsequent
washout of the baryon number by sphalerons is avoided.
Quantitatively this constraint translates to \cite{shap}
\begin{equation}
v(T_c)/T_c \stackrel{>}{\sim} 1, \label{const1}
\end{equation}
where $v(T_c)$ is the Higgs vacuum expectation value (VEV) at the
first order phase transition temperature $T_c$. Because $v(T_c)$ is
given in terms of the parameters of the Higgs potential, another
important constraint for any model that aims to satisfy Eq.~(\ref{const1})
is the experimental lower bound on the Higgs mass \cite{alep}:
\begin{equation}
m_H \stackrel{>}{\sim} 60 {\rm GeV}. \label{const2}
\end{equation}

So far, we have seen the following picture emerge from this topic of
effective potential construction in various models (the list below is
by no means exhaustive in this area):\newline
1) {\it The minimal SM}: A first order phase transition exists
due to temperature induced cubic terms in the effective potential,
but the strength does not appear to be enough and any baryon asymmetry
created is likely to be washed out after the phase transition - that is,
constraints (\ref{const1}) and (\ref{const2}) cannot be satisfied
simultaneously \cite{carr,buch}.\newline
2) {\it The SM + an extra Higgs doublet}: There seems to be a sufficient
number of parameters present to satisfy both constraints \cite{boch}.\newline
3) {\it The minimal supersymmetric model (MSSM)}: In a manner similar
to that in the SM, it appeared that constraints (\ref{const1}) and
(\ref{const2}) could not be satisfied simultaneously \cite{giud}, but a
recent paper that included the higher loop corrections through the
self-energies claims to have found a small parameter space where this
is possible \cite{espi}.\newline
4) {\it The MSSM + a Higgs singlet}: It appears that due to the presence of
tree-level cubic terms, constraints (\ref{const1}) and (\ref{const2})
can be satisfied even without help from the temperature induced
cubic terms \cite{piet}.\newline
5) {\it The left-right symmetric model}: The minimal Higgs sector which
consists of two Higgs doublets gives a similar result to the SM case
\cite{choi}, while the alternative Higgs sector with two triplets and a
bidoublet seems to require the presence of a further singlet Higgs for
sufficient baryon asymmetry generation \cite{moha}. \newline
6) {\it The SM + a complex singlet Higgs}: For the case in which the VEV
of the singlet remains zero for all temperatures it is possible to
satisfy both constraints (\ref{const1}) and (\ref{const2}) \cite{ande}.
In the singlet Majoron case, with the VEV of the singlet non-zero, it
appears that either (i) both constraints can be satisfied simultaneously
for somewhat large quartic couplings \cite{kond}; or (ii) that constraint
(\ref{const2}) may be modified through the presence of Majoron decay
channels for the physical Higgs boson hence keeping the viability of
this model alive \cite{enqv}.

In this paper, we consider a {\it real} Higgs singlet that couples only to
the standard Higgs doublet. This is the smallest possible extension of the
SM Higgs sector. The motivation for this work is twofold.
First, we want to contribute to the above picture which basically aims
to find those models compatible with electroweak baryogenesis and to try
to find perhaps some consistent pattern emerging amongst the possible
scenarios. Second, unlike the complex singlet case, we will
have a tree-level cubic term, rather like the case of the singlet
Higgs added to the MSSM. There, the tree-level cubic term appeared
to play a significant role in ensuring a strong first order phase transition.
Is this a promising trend to be found elsewhere too? We keep the singlet
VEV non-zero thus ensuring the generality of our analysis. We also do not
impose any specific restriction on the parameters, working numerically
with a random selection of parameter sets to obtain results.

\section{The effective potential}
We add a real singlet Higgs field $S$ to the minimal SM with its doublet
Higgs field $\phi$. As usual, $\phi\sim\underline{2}(1)$ under
${\rm SU(2)}_L\otimes{\rm U(1)}_Y$ while $S\sim\underline{1}(0)$.
Hence the most general potential at zero temperature is:
\begin{equation}
 V_o(\phi, S) = \lambda_{\phi}(\phi^{\dagger}\phi)^2
                -\mu_{\phi}^2\phi^{\dagger}\phi
                + \frac{\lambda_S}{2}S^4 - \frac{\mu_S^2}{2}S^2
                -\frac{\alpha}{3}S^3 + 2\lambda(\phi^{\dagger}\phi)S^2
                - \frac{\sigma}{2}(\phi^{\dagger}\phi)S.
\label{Vo}\end{equation}
Expanding each Higgs field about constant background fields, we
obtain the tree-level potential,
\begin{equation}
 V_{\rm tree}(u,v) = \frac{\lambda_{\phi}}{4}u^4 -\frac{\mu_{\phi}^2}{2}u^2
                   + \frac{\lambda_S}{4}v^4- \frac{\mu_S^2}{2}v^2
                   -\frac{\alpha}{3}v^3 + \frac{\lambda}{2}u^2v^2
                   - \frac{\sigma}{4}u^2v,
\label{Vtree}\end{equation}
where $\quad \phi(x) = \frac{1}{\sqrt{2}}[\left(\begin{array}{c}
0\\u \end{array}\right)+\Phi(x)]\quad\mbox{and}\quad
S(x) = v + \eta(x)$.
Note that $V_{\rm tree}(u,v)$ is symmetric in $u\rightarrow -u$ but not
in $v\rightarrow -v$. Hence we will restrict our attention to the
$u\geq 0$ section of the $(u,v)$ plane. Setting $\nabla V_{\rm tree}
(u,v)=0$ gives the following equations for the points of zero gradient:
\begin{equation}\begin{array}{l}
(-\mu_{\phi}^2 + \lambda_{\phi}u^2 + \lambda v^2 - \sigma v/2)u=0\\
-\mu_S^2 v + \lambda_S v^3 + \lambda v u^2 - \sigma u^2/4 - \alpha v^2=0.
\end{array}\label{mineq}\end{equation}
There are up to six different solutions to Eq.~(\ref{mineq}) in $u\geq 0$
part of the $(u,v)$ plane. Some of these are saddle points, while others
are local minima or maxima. Up to two local minima can exist, with
the only constraint on these being that the global minimum
$\equiv(\kappa_1,\kappa_2)$ obeys $\kappa_1 = 246$ GeV.

The field dependent Higgs masses are given by,
\begin{eqnarray}
  m_{\phi,S}^2(u,v) = \frac{1}{2}\left[ -\mu_{\phi}^2-\mu_S^2+
    (3\lambda_{\phi}+\lambda)u^2 +(3\lambda_S+\lambda)v^2 -
    (\frac{\sigma}{2}+2\alpha)v \right]\qquad\qquad\quad\qquad\nonumber\\
   \pm\left\{\frac{1}{4}
     \left[ -\mu_{\phi}^2+\mu_S^2+(3\lambda_{\phi}-\lambda)u^2
            -(3\lambda_S-\lambda)v^2 - (\frac{\sigma}{2}-2\alpha)v \right]^2
     +(2\lambda v - \frac{\sigma}{2})^2u^2\right\}^{1/2}\nonumber\\
 m_g^2(u,v) = \lambda_{\phi} u^2 + \lambda v^2 -\sigma v/2 - \mu_{\phi}^2,
\qquad\qquad\qquad\qquad\qquad\qquad\qquad\qquad\qquad\qquad\quad
\label{Higgs mass}\end{eqnarray}
where $m_g^2(u,v)$ is the contribution from goldstone bosons.
Because $S$ does not couple to the gauge bosons or fermions, their field
dependent masses are the same as in the minimal SM:
\begin{equation}
 m_W^2(u) = g^2u^2/4, \quad m_Z^2(u) = (g^2 + g^{'2})u^2/4,
 \quad m_A^2(u) = 0, \quad m_t^2(u) = Y^2u^2/2,
\label{g.b. mass}\end{equation}
where $Y$ is the top Yukawa coupling constant.

Using standard methods, we obtain the 1-loop effective potential \cite{dola}
\begin{equation}
 V(u,v) = V_{\rm tree}(u,v) + V_1^{(0)}(u,v) + V_1^{(T)}(u,v)
\end{equation} where
\begin{equation}
 V_1^{(0)}(u,v) = \sum_i \frac{n_i}{64\pi^2} m_i^4(u,v)
           \left[ {\rm ln}\frac{m_i^2(u,v)}{m_{0i}^2} - \frac{3}{2}\right]
\end{equation} and
\begin{equation}
 V_1^{(T)}(u,v) = \sum_i \frac{n_i}{2\pi^2} T^4
                 I_{\pm}\left( \frac{m_i^2(u,v)}{T^2} \right).
\end{equation}
The sum is over all the particles in the model, with $m_i^2(u,v)$ as given
in Eqs.~(\ref{Higgs mass}) and (\ref{g.b. mass}). The quantities
$n_i$ account for the degrees of freedom of each particle,
and $m_{0i}$ is the mass of the particle at zero temperature. The functions
$I_{\pm}$ are given by
\begin{equation}
 I_{\pm}(y) = \int_0^{\infty} dx\,x^2\,{\rm ln}(1\mp e^{\sqrt{x^2+y}}),
\end{equation}
where $I_+$ is used for bosons and $I_-$ for fermions.
This one loop result can be improved by summing up the dominant higher
order loop corrections, a procedure that introduces the self-energies
$\Pi_i(0)$ for the appropriate particle degrees of freedom \cite{dola,gros}.
By including only the leading high temperature correction and expanding
$I_{\pm}$ in powers of $m_i/T$, we obtain
\begin{equation}
 V_1^{(T)}(u,v)=\sum_i n_i\left[\frac{m_i^2(u,v)T^2}{24}
    -\frac{3}{2}\frac{m_i^4(u,v)}{64\pi^2}\right]
    -\sum_j n_j\frac{M_j^3(u,v) T}{12\pi},
\end{equation}
where $M_j^2(u,v) = m_j^2(u,v) + \Pi_j(0)$ for $j=g,W_L,W_T,Z_T,\gamma_T$
[where subscripts $L$ and $T$ refer to longitudinal and transverse modes
respectively], and for $j=\phi,S$, one can use the $M_j^2(u,v)$ as the
diagonal elements of the initial mass matrix modified by the addition
of the self-energies of the Higgs bosons.

This finite temperature effective potential (FTEP) is quite similar to
that of Enqvist {\it et.al.} \cite{enqv} except that we have the
additional terms $-\frac{\alpha}{3}v^3 -\frac{\sigma}{4}u^2v$. They
found a weak first order phase transition in the $u=0$ direction followed
by another first order phase transition in the $v\simeq$constant direction
induced by finite temperature cubic terms. Clearly the same scenario should
be viable in this model also, as we have extra parameters $\alpha$ and
$\sigma$. But the presence of these parameters should give us an
extra effect. Because no Majoron is present in this model, we cannot
modify constraint~(\ref{const2}) as was done by Enqvist {\it et.al.}
So we must rely on these extra parameters to allow us to satisfy
both of the constraints.

\section{Analysis of the finite temperature effective potential}
As a first step, we omit the temperature induced cubic terms altogether.
This will give us a clear indication of the magnitude of the role
played by the $\alpha$ and $\sigma$ terms, since no first order phase
transition can exist without the temperature induced cubic terms
within the minimal SM or the singlet Majoron model. The temperature
independent 1 loop potential terms will be neglected for simplicity
(they only contribute a few percent to the overall potential in the region
$m/T<1$). Now the FTEP can be written down explicitly:
\begin{equation}
V(u,v)=V_{\rm tree}(u,v) + T^2\left[ (\frac{6\lambda_{\phi}+\lambda}{24}+
       0.11)u^2 + \frac{3\lambda_S+4\lambda}{24}v^2
       - \frac{\sigma+\alpha}{12}v \right],
\end{equation}
where we have used $g=0.652$, $g^{'}=0.352$ and $Y=0.73$ ($\Rightarrow
m_t= $127 GeV; this choice for $Y$ is illustrative only, and corresponds
to the most likely value for the top mass within the context of the minimal
SM).

There are seven free parameters $(\lambda_{\phi},\lambda_S,\lambda,
\mu_{\phi}^2,\mu_S^2,\alpha,\sigma)$, with one constraint that the
global minimum at zero temperature is set at $u=246$ GeV. To take this
into account more easily, we express $(\mu_{\phi}^2,\mu_S^2,\alpha,\sigma)$
in terms of $(u_1,v_1,u_2,v_2)$, the values of VEV's at which the gradient
of $V_{\rm tree}(u,v)$ is zero. We then choose $u_1=246$ GeV at zero
temperature. We also choose to concentrate on those cases where
$v_1=-v_2$ which makes the pattern of VEV formation described in Figs.~1
to 4 more realisable. This leaves us with 5 parameters:
$(\lambda_{\phi},\lambda_S,\lambda,u_2,v_2)$.
We then choose random sets of these in the following ranges:
\begin{equation}
10^{-3}<\lambda_{\phi},\lambda_S<10^{-1},\qquad -0.5<\lambda<0.5,\qquad
100 {\rm GeV}<u_2,v_2<300 {\rm GeV}.
\end{equation}
The range in $\lambda_{\phi},\lambda_S$ is chosen so that $V(u,v)$
which comes partly from expanding in these parameters makes sense.
The range in $\lambda$ is chosen to be general, with one constraint
that $\frac{\lambda_{\phi}}{4}+\frac{\lambda_S}{4}+\frac{\lambda}{2}>0$
to ensure that $V_{\rm tree}$ is bounded below along $u=v$ line.
The VEV's $u_2,v_2$ are essentially unconstrained and we have simply
chosen them to be at the same scale as $u_1$.

We then went through these randomly chosen sets of parameters and
tested each set for the following conditions, rejecting the set if any
of these conditions was not met:\newline
\begin{equation}\begin{array}{l}
\mbox{1) The global minimum at zero temperature is at} (u_1,v_1),
\mbox{ where }u_1=246 \mbox{ GeV.}\\
\mbox{2) The mass of the lightest Higgs boson} \geq 60 \mbox{ GeV
[constraint~(\ref {const2})].}\\
\mbox{3) A first order phase transition exists for some temperature } T_{c}.\\
\mbox{4) }m_i/T_c < 1 \quad\mbox{at the points in the }(u,v)\mbox{ plane
 where a first order phase}\\
\quad\mbox{transition occurs.}
\end{array}\label{condns}\end{equation}

Figs.~1 to 4 show the most promising pattern of VEV formation (as
temperature changes) that we found in terms of satisfying both
constraints (\ref{const1}) and (\ref{const2}). (The second set of
parameters in Table~1 is used to generate these figures.) Above the
transition temperature the global minimum lies at $(U_1^T,V_1^T)$ where
$U_1^T>0$ and $V_1^T<0$. At the temperature $T_c$ (which is of order 100
GeV), two degenerate minima form at points $(U_1^{T_c},V_1^{T_c})$ and
$(0,V_2^{T_c})$, where $V_2^{T_c}>0$, and there is a barrier between
the two minima. A first order phase transition occurs as the global
minimum ``tunnels'' from $(U_1^{T_c},V_1^{T_c})$ point to $(0,V_2^{T_c})$
point. Table~1 shows some of the allowed sets of parameters and the values
of some relevant quantities. It reveals how constaints (\ref{const1})
and (\ref{const2}) can both be met.
In general we find that as $T\rightarrow \infty$, the global minimum
$\rightarrow (0,V_{\infty})$, where $V_{\infty}\neq 0$. If we set
$\alpha = - \sigma$, $V_{\infty}=0$ is achieved. However, we find that
within the allowed sets of parameters, $V_{\infty}$ is small
[$\sim O(10)$]. At any rate because the electroweak-breaking VEV
$U_1^T$ has jumped to zero at $T_c$, electroweak symmetry
restoration does take place.

Qualitatively, the reason for this FTEP to allow such a strong first
order phase transition lies in the asymmetry in $v\rightarrow -v$.
Two valleys, roughly parallel to the $u-$axis, form as $T$ nears $T_c$,
in such a way as to allow two degenerate minima. In this regard the
tree level cubic terms $\alpha$ and $\sigma$ are essential.
Note that for most of the sets of parameters in Table~1, we find that in
addition to the usual electroweak phase transitions, another first order
phase transition purely in $v$ direction can occur, at temperatures
higher than the electroweak phase transition temperature (see the caption
for Fig.~3). However this additional first order phase transition is not
a requirement for successful baryogenesis and the last set of
parameters in Table~1 is an example where only one first order phase
transition occurs (for this set, $v_2=200$ GeV).

\begin{table}
\caption{Representative parameter values for which the conditions (15)
are met. $V_c$ is the distance between the two degenerate
minima in the $(u,v)$ plane at $T_c$. The parameters
$\lambda_{\phi,S}$ and $\lambda$
are dimensionless while all the other quantities have dimension GeV.}
\[\begin{array}{cccccccc}
  \lambda_{\phi}\quad & \lambda_S\quad & \lambda\quad & v_1\quad
  & \sigma\quad & \alpha\quad & m_{\phi,S}\quad & V_c/T_c\quad \\ \hline
0.071\quad & 0.082\quad & -0.0234\quad & -125\quad & -12.81\quad &
            12.20\quad & 61,102\quad & 170/107\\
0.063\quad & 0.073\quad & 0.0250\quad & -154\quad & -13.26\quad &
            3.58\quad & 73,88\quad & 390/93\\
0.087\quad & 0.090\quad & 0.0114\quad & -245\quad & -9.37\quad &
            1.24\quad & 103,108\quad & 510/117\\
0.096\quad & 0.082\quad & 0.0499\quad & -146\quad & -17.27\quad &
            5.10\quad & 75,109\quad & 430/97\\
0.050\quad & 0.077\quad & 0.0168\quad & -117\quad & -12.43\quad &
            8.34\quad & 67,79\quad & 360/70\\
0.057\quad & 0.0920\quad & 0.0026\quad & -171\quad & -16.39\quad &
            4.64\quad & 73,95\quad & 360/85\\
0.057\quad & 0.0920\quad & 0.0026\quad & -107\quad & -16.54\quad &
            16.76\quad & 68,92\quad & 360/63
\end{array}\]
\end{table}

What would happen if we were to include the temperature dependent
cubic terms? Adding the gauge boson cubic terms $\sim g^3T(u^2)^{3/2}$
may change the shape of the potential along the valleys but not the
degenerate nature of the two valleys themselves as these terms are
independent of $v$. The Higgs boson cubic terms $\sim \frac{T}{12\pi}
[(3\lambda_S+\lambda)v^2 + ... + c_i^2T^2]^{3/2}$, where the $c_i^2T^2$
piece comes from self-energy effects, are much more complicated and they
will depend on $v$. Within the minimal SM, including the self-energy pieces
in the Higgs boson cubic terms introduces an upper bound on the Higgs
boson mass, above which the phase transition becomes second order \cite{buch}.
The existence of this upper bound [in addition to the lower bound on the
Higgs mass that comes from satisfying constraint (\ref{const1})] further
restricts the available parameter space. Therefore one may wonder if the
sets of allowed parameters we found here may actually become excluded as
soon as the Higgs self-energy terms are included. But we are again saved
by the tree-level cubic terms, as the upper bound restriction no longer
applies when terms of the form $\alpha v^3$ exist (for any non-zero value
of $\alpha$). Hence adding the Higgs self-energy terms will not make the
first order phase transition go away. It will of course change the actual
value of the allowed parameters. To get some idea on the quantitative
effect on the parameters when these Higgs boson cubic terms are
included, we fitted cubic polynomials in $v$ to these terms (around
the regions of degenerate minima at $T_c$). The result is consistent
with what one may expect with a naive estimate of $\sim\frac{T}{12\pi}
(3\lambda_S+\lambda)^{3/2}v^3$. For the values of the parameters
shown in Table~1, there is at most an additional term of $\sim -0.3v^3$
coming from the Higgs boson cubic terms. This means that these terms
make less contribution for larger values of $\alpha$ (for example,
when $\alpha=12.20$, $-0.3v^3$ term is about 7\% of the tree-level
cubic term $-\alpha v^3/3$). We thus conclude that our qualitative
conclusions obtained without inclusion of the temperature-dependent
cubic terms are robust, and that our approximate effective potential
is quantitatively accurate to about the 10\% level near the degenerate
minima at temperatures around $T_c$.

\section{Conclusion}

We have investigated the minimal extension of the SM Higgs sector
at finite temperature by writing down the effective potential when
a real singlet Higgs boson is included. We found that one can find
sets of parameters such that a strong first order phase transition
that will prevent a washout of possible baryon asymmetry can exist,
while satisfying the lower bound on the experimentally observed Higgs
mass. Our main point is that by having tree-level cubic terms in
the Higgs potential these conditions are more easily satisfied. When
one considers the difficulties faced by models that rely solely on
the temperature dependent cubic terms in satisfying these constraints,
tree-level cubic terms provide perhaps a good rough measure of the
potential viability of the models for electroweak baryogenesis.
\footnote{We note that a first-order electroweak phase transition is
a necessary but not sufficient condition to generate adequate
baryon asymmetry. Other issues such as adequate CP-violation and
appropriate bubble-wall propagation are also important, but these
are beyond the scope of our present analysis.}

\section*{Acknowledgments}
The authors would like to thank the following people for useful
discussions: D. Burke, J. Daicic, C. Dettmann, R. Gailis and
B. Hanlon. One of us (J.C.) would like to acknowledge the support of the
Australian Postgraduate Research Program. R. R. V. is
supported by grants from the University of Melbourne.

\section*{Figure captions}
\begin{itemize}
\item[Fig.1:]
A contour plot of $V(u,v)$ at $T=0$. The position of global minimum is
denoted by the dot at $(u,v)=(246,-154)$ (all units are in GeV). There
are saddle points at $(0,-162)$ and $(0,211)$. There is a local minimum
at $(168,154)$ which will move towards $(0,200)$ in a second order phase
transition as $T$ increases. (The second set of parameters in Table~1 is
used to generate these figures.)
\item[Fig.2:]
A contour plot of $V(u,v)$ at the transition temperature, 93 GeV.
The positions of degenerate global minima are denoted by the dots
at $(0, 200)$ and $(160,-152)$.
\item[Fig.3:]
A contour plot of $V(u,v)$ at $T=200$ GeV. The local minimum at
$(160,-152)$ at $T=93$ has moved to $(0,-130)$ in a second order
phase transition. After this, another first order phase transition
has occurred from $(0,156)$ to $(0,-130)$, making the latter the global
minimum now. This extra first order phase transition is however not
necessary for adequate baryogenesis.
\item[Fig.4:]
A contour plot of $V(u,v)$ at $T=400$ GeV. The global minimum is now
at $(0,-58)$. As $T$ increases further, this global minimum will settle
towards $(0,-30)$.
\end{itemize}


\begin{thebibliography}{99}
\bibitem{nels}For a review, see {\em Progress in electroweak baryogenesis}
              by A.G.~Cohen, D.B.~Kaplan and A.E.~Nelson, UCSD-93-2,
              BU-HEP-93-4 (to appear in Annual Review of Nuclear and
              Particle Science, Vol. 43), and the references therein.
\bibitem{shap}M.E.~Shaposhnikov, JETP Lett. 44 (1986) 465;
              Nucl. Phys. B287 (1987) 757; B299 (1988) 797; Phys. Lett. B277
              (1992) 324, Erratum, Phys. Lett. B282 (1992) 483.
\bibitem{alep}ALEPH, DELPHI, L3 and OPAL collaborations, as presented
              by M. Davier, Proceedings of the International Lepton-Photon
              Symposium and Europhysics Conference on High Energy Physics,
              eds. S. Hegerty, K. Potter and E. Quercigh (World Scientific,
              1992, Singapore).
\bibitem{carr}M.E.~Carrington, Phys. Rev. D45 (1992) 2933;
              M.~Dine, R.G.~Leigh, P.~Huet, A.~Linde and D.~Linde,
              Phys. Lett. B283 (1992) 319 and Phys. Rev. D46 (1992) 550;
              P.~Arnold, Phys. Rev. D46 (1992) 2628;
              J.R.~Espinosa, M.~Quir\'os and F.~Zwirner, Phys. Lett. B291
              (1992) 115;
              C.G.~Boyd, D.E.~Brahm and S.D.~Hsu, Caltech preprint
              CALT-68-1795, HUTP-92-A027, EFI-92-22;
              W.~Buchm\"uller and T.~Helbig, preprint DESY 92-117;
              W.~Buchm\"uller, T.~Helbig and D.~Walliser, preprint
              DESY 92-151;
              P.~Arnold and O.~Espinosa, Phys. Rev. D47 (1993) 3546.
\bibitem{boch}A.I.~Bocharev, S.V.~Kuzmin and M.E.~Shaposhnikov,
              Phys. Lett. B244 (1990) 275, Phys. Rev. D43 (1991) 369;
              N.~Turok and J.~Zadrozny, Nucl. Phys. B358 (1991) 471
              and Nucl. Phys. B369 (1992) 729;
              B.~Kastening, R.D.~Peccei and X.~Zhang, Phys. Lett. B266 (1991)
              413; L.~McLerran {\it et.al.}, Phys. Lett. B256 (1991) 451.
\bibitem{giud}G.F.~Giudice, Phys. Rev. D45 (1992) 3177;
              S.~Myint, Phys. Lett. B287 (1992) 325.
\bibitem{espi}J.R.~Espinosa, M.~Quir\'os and F.~Zwirner, Phys. Lett. B307
              (1993) 106.
\bibitem{piet}M.~Pietroni, Padova preprint DFPD/92/TH/36.
\bibitem{choi}J.~Choi and R.R.~Volkas, University of Melbourne preprint
              UM-P-92/88, [Phys. Rev. D, in press (1993)].
\bibitem{moha}R.N.~Mohapatra and X.~Zhang, Phys. Rev. D46 (1992) 5331.
\bibitem{ande}G.W.~Anderson and L.J.~Hall, Phys. Rev. D45 (1992) 2685;
              J.R.~Espinosa and M.~Quir\'os, Phys. Lett. B305 (1993) 98.
\bibitem{kond}Y.~Kondo, I.~Umemura and K.~Yamamoto, Phys. Lett. B263
              (1991) 93; N.~Sei, I.~Umemura and K.~Yamamoto, Phys. Lett.
              B299 (1993) 286.
\bibitem{enqv}K.~Enqvist, K.~Kainulainen and I.~Vilja, preprint
              NORDITA-92/10P (1992).
\bibitem{dola}L.~Dolan and R.~Jakiw, Phys. Rev. D9 (1974) 3320;
              S. Weinberg, Phys. Rev. D9 (1974) 3357;
              D.A. Kirzhnits and A.D. Linde, Sov. Phys. JETP 40 (1975) 628,
              Ann. Phys. 101 (1976) 195.
\bibitem{gros}D.J. Gross, R.D. Pisarski and L.G. Yaffe, Rev. Mod. Phys.
              53 (1981) 43;
              P. Fendley, Phys. Lett. B196 (1987) 175;
              J.I. Kapusta, {\em Finite temperature Field Theory} (Cambridge
              University Press, 1989).
\bibitem{buch}J.R.~Espinosa, M.~Quir\'os and F.~Zwirner, preprint
              CERN-TH.6577/92, IEM-FT-58/92;
              W.~Buchm\"uller, Z.~Fodor, R.~Helbig and D.~Walliser, preprint
              DESY 93-021.
\end{thebibliography}
\end{document}